# ESTUDIO DEL PROCESO DE ADSORCIÓN-DESORCIÓN DE CONTAMINANTES EN MEDIOS CONFINADOS MEDIANTE SIMULACIONES COMPUTACIONALES

Estela MAYORAL[1], Eduardo DE LA CRUZ[1],
Luis Carlos LONGORIA[1] y Eduardo NAHMAD-ACHAR[2]

[1] Instituto Nacional de Investigaciones Nucleares, Carretera México-Toluca s/n, La Marquesa Ocoyoacac, Estado de México CP 52750, México. Correos electronicos: estela.mayoral@inin.gob.mx, eduardo.delacruz@inin.gob.mx
[2] Instituto de Ciencias Nucleares, Universidad Nacional Autónoma de México (UNAM), Apartado Postal 70-543, 04510 México D.F. Correos electrónicos: luis.longoria@inin.gob.mx, nahmad@nucleares.unam.mx

## RESUMEN

El estudio de la dispersión de contaminantes en medios porosos es de suma importancia para estimar el impacto ecológico que los desechos sólidos de origen radioactivo ocasionan. Existen modelos que simulan la difusión de soluto en medios porosos considerando las características físico-químicas del contaminante disperso y su efecto en la movilidad, la adsorción-desadsorción y la adsorción irreversible sobre las paredes del canal a través del cual se difunde. La mayoría de estos modelos no considera la presencia de distintas especies interaccionando unas con otras y con el sustrato y las consecuencias de esta competencia entre distintas especies en el equilibrio termodinámico del sistema. En este trabajo se presenta el estudio del proceso de adsorción-desorción de distintos componentes presentes en los fluidos contaminados a través de simulaciones moleculares mesoscópicas electrostáticas tipo DPD (dissipative particle dynamics). Se muestra cómo la presencia de distintos componentes altera el equilibrio de adsorción-desorción, indicando que las simulaciones de dispersión del contaminante deberán tomar en cuenta, este cambio en la constante de adsorción para obtener resultados más cercanos a la realidad.

## ABSTRACT

Nowadays, the study of dispersion of solid wastes through a porous media is very important in order to estimate the ecological impact that, in particular a radioactive solid waste, could produce when it spreads in the soil. There are some models available in literature which can help us to simulate the dispersion of contaminants through a porous media, taking into account the physicochemical properties of the waste and its effect over the mobility, the adsorption-desorption equilibrium and the irreversible adsorption over the walls that constitute the channel where it diffuses. However, the majority of these models do not consider the cooperative behavior given by the presence of other species competing each other for the substrate and the consequences that this competition produces in the thermodynamic equilibrium of the system. In this sense, the mesoscopic simulations have shown to be a viable alternative to study these kinds of systems. This work presents the study of the adsorption-desorption process for different components in a contaminated fluid by electrostatic mesoscopic molecular simulation using the dissipative particle dynamics method (DPD). Also it shows how the presence of different components modifies the adsorption-desorption equilibrium, this result suggests the importance of including, in the simulation of contaminated fluid in a porous media, the presence of all species in order to obtain results closer to reality.

---

## INTRODUCCIÓN

El estudio de la dispersión de contaminantes muy nocivos a través del suelo es de gran relevancia, debido a la necesidad de estimar el impacto ecológico a largo plazo que cierto tipo de contaminantes puede ocasionar. Junto con los desechos industriales muy tóxicos, específicamente la contaminación por desechos radioactivos resulta ser de primordial importancia dada su peligrosidad. Debido a que la etapa final en el tratamiento de este tipo de desechos es su confinamiento en lugares con características específicas que aseguren su aislamiento hasta decaer sin provocar daño a los seres vivos, se requiere de una serie de estudios para hacer la selección del lugar más adecuado para dicho propósito.

Con base en las normas de seguridad establecidas (Safety Guide, No. WS-G-1.1 1999) la selección del sitio de confinamiento debe considerar aspectos tanto radiológicos como ambientales, así mismo, debe de llevarse a cabo el estudio mediante modelos que simulen la evolución del sistema e indiquen las consecuencias posibles en distintos escenarios. Dichos modelos deben estar calibrados y validados de tal manera que reproduzcan suficientemente bien el sistema real. Para este propósito se requiere del conocimiento del conjunto de parámetros de entrada propios del sistema a simular.

Actualmente, este tipo de simulación se hace con modelos matemáticos macroscópicos que se basan en la ecuación de difusión y transporte de soluto en medios porosos saturados (Fetter 1994), tomando en cuenta primordialmente tres componentes que dependen de las características fisicoquímicas del contaminante disperso en el medio móvil. Estas son la movilidad (debida a la difusión en el medio), los procesos de adsorción-desorción y la fijación del contaminante en forma irreversible a las paredes del canal (esto es, procesos de quimiosorción) (Bear 1979). Además de considerar propiedades características como el decaimiento y la actividad del radionúclido, esta última considerada como la magnitud que expresa la velocidad de transformación de los núcleos radiactivos y que es proporcional al número de átomos radiactivos $N$, esto es $A=\lambda N$ donde $\lambda$ es la constante de desintegración radiactiva y es característica de cada radionúclido (Iturbe 2001). Para una predicción más cercana a la realidad, el éxito del modelo macroscópico depende de los parámetros de entrada que se le proporcionen. Estos deben ser obtenidos experimentalmente o bien estimarlos considerando su dependencia con distintas condiciones del medio, tomando en cuenta por ejemplo pH, salinidad, reacciones químicas y fisicoquímicas alternas, temperatura y presión. Dada la sensible dependencia de estas variables con el medio, el adecuado conocimiento de estos parámetros permite predecir el comportamiento del sistema a tiempos largos, el cual estará determinado por el tiempo de vida media del radionúclido y bajo las condiciones del medio donde se encuentren inmersos en la realidad.

Debido a la elevada peligrosidad de los componentes radiactivos, la obtención experimental de estas cantidades representa por sí misma un riesgo, inevitablemente se generan residuos por el propio procedimiento y en muchos casos son muy costosos o prácticamente imposibles de realizar. Para evitar este riesgo, los métodos de modelado molecular han mostrado ser una alternativa limpia y segura empleando métodos de dinámica molecular microscópica, Monte Carlo y simulaciones cuánticas (Fermeglia y Pricl 2007). Estas simulaciones hacen posible estimar parámetros fisicoquímicos, termodinámicos y estructurales (Frenkel y Smit 2002) en condiciones difíciles de obtener experimentalmente, así como obtener valores adecuados para las constantes de adsorción-desorción y estudiar el proceso dinámico del fenómeno competitivo en la adsorción, fundamental cuando se encuentran presentes varias especies en el medio. Con esta información es posible obtener los parámetros adecuados que caractericen la fase móvil, la sorbida y la fijada e introducirlos en el modelo macroscópico.

Una de las principales fuentes de desechos radiactivos que requieren ser almacenadas en forma segura en depósitos geológicos, son los desechos de $^{238}U$ y $^{235}U$, principales componentes del combustible nuclear, que después de su uso y desgaste se convierten en desechos radiactivos de alto nivel. Su tiempo de vida media es del orden de miles de millones de años por lo tanto es necesario considerar la posibilidad de alguna fuga por eventos extraordinarios, y en consecuencia estimar la migración de este contaminante a través del subsuelo. El uranio puede presentarse en distintas formas, siendo el $U(IV)$ la forma termodinámica más estable de este elemento. Al entrar en contacto con un medio acuoso forma una cantidad importante de especies iónicas de las cuales destaca el $UO_2^{2+}$ (uranilo) y sus formas hidrolizables. Este compuesto tiene la característica de ser altamente móvil y de interaccionar fuertemente con los componentes de las fases sólidas y en disolución formando diversos complejos cuya estabilidad depende principalmente del pH y del potencial redox del medio, este comportamiento define su transporte en el medio geológico alterando el proceso de difusión.

Debido a lo anterior, en este trabajo se presentan los resultados del estudio de la adsorción competitiva del contaminante iónico radiactivo (uranilo $UO_2^{2+}$) en presencia de componentes orgánicos típicos del suelo. Para estos últimos se consideraron aquellos que provienen de la desintegración natural de los aminoácidos presentes en la materia biológica como por ejemplo grupos carboxilatos en forma de sales de ácido poliacrílico. También se consideran compuestos externos derivados de procesos, desechos industriales y de consumo común como glicoles, en medios confinados para estimar las concentraciones de saturación de dichas especies sobre sustratos representativos del suelo como óxidos metálicos mediante simulaciones mesoscópicas de dinámica de partículas disipativas electrostáticas (DPD).

Este método de simulación mesoscópica es una novedosa alternativa limpia y económica que permite estudiar medios que involucran componentes de alta peligrosidad, evitando la innecesaria exposición a ciertos materiales radiotóxicos, conservando el apego a los principios básicos de protección radiológica [NOM-022/1-NUCL-1996]. Aunado a ello permite superar una de las principales limitantes de la simulación molecular en cuanto a la gran cantidad de recursos computacionales que se requieren para estudiar problemas de sistemas complejos como la adsorción competitiva de distintos componentes en sustratos naturales En este sentido las simulaciones DPD han mostrado ser una alternativa viable para estudiar este tipo de sistemas (Hoogerbrugge y Koelman 1992). Aunado a ello, los resultados obtenidos de este tipo de simulaciones permiten retroalimentar de manera más rápida los modelos macroscópicos utilizados para estudiar la migración de contaminantes en diferentes medios.

**Modelos de dispersión en medios porosos**

Para simular el transporte de soluto en medios porosos saturados, existen modelos matemáticos macroscópicos que, en general, consideran tres componentes relacionadas con el material dispersado en suelo u otro sustrato poroso (Fetter 1994, Bear 1979): la móvil ($m(z,t)$), la sorbida ($s(z,t)$) y la ligada ($b(z,t)$) para evaluar el cambio en concentración de dicho componente durante el tiempo. La componente móvil del radioisótopo se define como la que puede migrar por la presencia de un gradiente de concentración. La componente sorbida es la que se encuentra ligada en modo reversible a la superficie del suelo mientras que, la componente fija, se encuentra ligada en forma irreversible. Si definimos como concentración libre a $f(z,t) \equiv m(z,t) + s(z,t)$, la concentración total del radioisótopo en suelo será $c(z,t) = f(z,t) + b(z,t)$. Si se toma en cuenta que la velocidad del proceso de fijación es proporcional a la componente móvil, se puede reescribir la difusión con base en la Ley de Fick (Toso y Velasco 2001) de la siguiente forma:

$$\frac{\partial f(z,t)}{\partial t} = D \frac{\partial^2 m(z,t)}{\partial^2 t} - km(z,t) - \lambda f(z,t)$$

$$\frac{\partial b(z,t)}{\partial t} = km(z,t) - \lambda b(z,t)$$

donde $D(cm^2 d^{-1})$ es el coeficiente de difusión para la componente móvil, $k(d^{-1})$ es la tasa de transferencia de la componente móvil a la ligada y $\lambda(d^{-1})$ es la constante física de decaimiento de algún radioisótopo. Las constantes $D$ y $k$ dependen de las características del suelo y $\lambda$ del contaminante. Para éste modelo macroscópico o algún otro que se desee emplear uno de los parámetros fundamentales que se requiere conocer es la capacidad de adsorción y retención de los componentes presentes. Por ello, en este trabajo el objetivo principal es el estudio de las propiedades de adsorción-desorción de iones uranilo sobre sustratos de óxidos metálicos en presencia de componentes naturales del suelo. Entre ellos se consideraron los derivados de la desintegración de aminoácidos prove-nientes de los seres vivos (compuestos orgánicos con grupos carboxílicos) y en presencia de contaminantes que pudieran provenir de actividad industrial (por ejemplo, el polietilenglicol).

## Simulaciones mesoscópicas: dinámica de partículas disipativas electrostática

Una de las limitantes en el área de simulación es el elevado costo computacional cuando se estudian fluidos complejos (sistemas constituidos por una gran cantidad de componentes interactuando a diferentes escalas de tiempo y tamaño). Aunado a la dinámica molecular y métodos Monte Carlo convencionales, el uso de simulaciones mesoscópicas basadas en la técnica de dinámica de partículas disipativas (DPD) ha permitido la obtención de resultados importantes (Español 2002). El modelo de Dinámica de Partículas Disipativas (Groot y Warren 1997), consiste en un conjunto de partículas (partículas DPD) que se mueven e interactúan entre ellas a través de una serie de fuerzas. Estas fuerzas son de tres tipos: una puramente repulsiva representada por la fuerza conservadora derivada de un potencial de energía y en la cual se conserva la energía, una fuerza disipadora que remueve energía del sistema e involucra el decremento de la velocidad relativa de las partículas y una fuerza aleatoria que actúa entre todos los pares de partículas y adiciona en promedio energía al sistema, junto con la fuerza disipativa actúa como un baño térmico es decir, que mantiene la temperatura del sistema constante. La característica distinguible de las fuerzas DPD es que se conservan el momento lineal, además de ser las responsables del comportamiento hidrodinámico de un fluido a gran escala. Desde el punto de vista físico, cada partícula disipativa no es considerada como un simple átomo de una molécula, sino como una colección de átomos que se mueven de manera coherente, contenidas dentro de una región o espacio más grande. Otra de las características atractivas de esta técnica es su enorme versatilidad para construir modelos simples de fluidos complejos ya que permite mapear varios átomos o moléculas del sistema real en una sola partícula DPD. Las ecuaciones básicas de DPD se describen a continuación.

Considérese un conjunto de partículas que interactúan donde la fuerza total entre las partículas $i$ y $j$, como se ha mencionado es la suma de una fuerza conservadora $\mathbf{F}_{ij}^C$, una fuerza disipadora $\mathbf{F}_{ij}^D$ y una fuerza aleatoria $\mathbf{F}_{ij}^R$, de tal manera que:

$$\mathbf{F}_{ij} = \mathbf{F}_{ij}^C + \mathbf{F}_{ij}^D + \mathbf{F}_{ij}^R \quad (1)$$

donde,

$$\mathbf{F}_{ij}^C = a_{ij}\,\omega^C(ij)\hat{e}_{ij}$$

$$\mathbf{F}_{ij}^D = \gamma\omega^D(r_{ij})[\hat{e}_{ij}\cdot\vec{v}_{ij}]\hat{e}_{ij}, \quad (2)$$

$$\mathbf{F}_{ij}^R = \sigma\omega^R(r_{ij})\hat{e}_{ij}\,\xi_{ij},$$

con vectores $\hat{e} = \mathbf{r}_{ij}\,/\,r_{ij}$, $\mathbf{r}_{ij} = \mathbf{r}_j - \mathbf{r}_i$, $\mathbf{v}_{ij} = \mathbf{v}_i - \mathbf{v}_j$, siendo $r_{ij}$ la distancia entre las partículas $i$ y $j$. Los vectores $\mathbf{r}_i$ y $\mathbf{v}_i$ son la posición y la velocidad de la partícula $i$, respectivamente. La fuerza aleatoria se calcula usando un número aleatorio $\xi_{ij}$ que está uniformemente distribuido entre 0 y 1, con media 0 y varianza 1. Las constantes $a_{ij}$, $\gamma$ y $\sigma$ determinan la intensidad de las fuerzas conservadora, disipadora y aleatoria respectivamente. Las funciones $\omega$, son funciones de peso dadas por

$$\omega^C(r_{ij}) = \omega^R(r_{ij}) = \sqrt{\omega^D(r_{ij})} = \omega(r_{ij}), \quad (3)$$

donde

$$\omega(r_{ij}) = \begin{cases} 1 - r_{ij}\,/\,R_c & r_{ij} \le R_c \\ 0 & r_{ij} \ge R_c \end{cases} . \quad (4)$$

La distancia $R_c$ es el radio de corte usado en la simulación. La fuerza conservadora transporta la contribución de las fuerzas que da lugar a la presión, la componente disipadora emula la viscosidad del medio y la fuerza aleatoria está asociada con el movimiento browniano de las partículas. La parte conservadora de la fuerza $\mathbf{F}_{ij}^C$ determinará el comportamiento termodinámico del sistema. Las fuerzas disipadora y aleatorias representan el termostato a temperatura $T$.

Las constantes $a_{ij}$ que caracterizan la interacción de cada partícula DPD se extraen a través de los parámetros de solubilidad (Español and Warren 1995) de los fragmentos monoméricos que representa cada cuenta DPD. Así, a través de estos parámetros de solubilidad que pueden ser obtenidos mediante simulaciones de dinámica molecular es posible mapear la información fisicoquímica de cada molécula.

Las interacciones electrostáticas son fundamentales en sistemas de interés con alto grado de especialización y Groot (2003) y González-Melchor *et al.* 2006) las introdujeron en el DPD. En ambos casos, la carga puntual en el centro de la partícula DPD se sustituye por una distribución de carga a lo largo de la partícula evitando así la formación de pares iónicos artificiales durante la simulación. González-Melchor *et al.* (2006), resuelven el problema adaptando el método estándar de Ewald (Toukmaji y Board 1996) para las partículas DPD mediante la técnica de sumas de Ewald (Toukmaji y Board 1996) que es la ruta más empleada para calcular las interacciones electrostáticas en simulaciones moleculares microscópicas.

## MATERIALES Y MÉTODOS

Debido a la complejidad de los sistemas originada por la presencia de grupos cargados con interacciones de distinto alcance y diferentes tamaños, se llevaron a cabo simulaciones mesoscópicas con el método DPD para sistemas electrostáticos empleando la metodología de sumas de Ewald descrito en la sección anterior y detallado en la referencia (González-Melchor *et al.* 2006). Con ello se simuló la dinámica de adsorción de los sistemas siguientes para estudiar y obtener las isotermas de adsorción de uno y dos componentes:

Sistema 1: $X_mO_n$ (sustrato) + $H_2O$ + $PAA^{N-}$ + $(N+r)$ $Na^+$ + $rCl^-$.
Sistema 2: $X_mO_n$ (sustrato) + $H_2O$ + PEG.
Sistema 3: $X_mO_n$ (sustrato) + $H_2O$ + PEG + $UO_2^{2+}$ + $Cl^{2-}$.
Sistema 4: $X_mO_n$ (sustrato) + $H_2O$ + $PAA^{N-}$ +N $Na^+$ + $UO_2^{2+}$ + $Cl^{2-}$.

El modelo de adsorción empleado considera dos placas paralelas genéricas representativas de óxidos metálicos $X_mO_n$, componentes presentes en el suelo las cuales actúan como sustrato. El fluido confinado entre estas dos placas paralelas fue $H_2O$ representadas en la simulación con 3 partículas DPD neutras y según el caso de estudio se incluyeron las siguientes partículas: un ión inorgánico radiactivo como el ión uranilo ($UO_2^{2+}$) representado por una cuenta DPD con carga $2^+$ y su correspondiente contraión ($Cl^{2-}$), un compuesto orgánico neutro como el polietilenglicol $[OH(CH_2CH_2O)_{n+1}H]$ que llamaremos PEG, representado por 7 cuentas DPD sin carga unidas por resortes y un compuesto orgánico polielectrolítico como lo es la sal de sodio de poliácido acrílico $(CH_3)_2[CH_2CH(COONa)]_N$ que denotaremos como $[PAA^{N-}][Na^+]_N$ representado por 20 cuentas DPD cada una con una carga de $1^-$ y unidas a través de resortes. A cada una le corresponde un peso molecular promedio de 1500 y 400 gr/mol, respectivamente. La **figura 1** muestra un esquema del mapeo de este sistema. Estos componentes fueron escogidos como prototipos representativos de contaminantes naturales e industriales de distinta peligrosidad para analizar el cambio en las propiedades de adsorción de los componentes cuando se encuentran presentes especies de distinta naturaleza.

Las simulaciones se llevaron a cabo empleando unidades adimensionales. Las masas son iguales a *1*, los valores de las constantes σ y γ fueron 3 y 4.5, respectivamente, por lo que $k_BT^*=1$. La caja de simulación fue de 9.8 X 9.8 X 24 unidades DPD (1 unidad DPD=6.46 Å) con 4300 partículas totales, de esta manera la densidad DPD del sistema (adimensional) es ρ*=3. El tiempo total de pasos de simulación fue de $n_T$=100 000 y el $\Delta t^*$ =0.02. Los polímeros fueron simulados con cuentas DPD representativas de cada grupo funcional unidas por resortes con una constante k=100. Los parámetros de interacción repulsivos $a_{ij}$ que corresponden a la fuerza conservadora empleados entre cada especie se obtuvieron a través de los parámetros de solubilidad de cada una de las especies (Español 2002, Esumi *et al.* 2001) y estos fueron: para la interacción $a_{PEG-H2O}$ =79.3, $a_{PAA-H2O}$ = 80 para las especies del mismo tipo y la interacción de los iones y el resto de las especies $a_{ii}$= 25, la interacción repulsiva entre las paredes y todas las especies fue de $a_{wi}$=200. El volumen total de cada partícula DPD fue de 90 Å³, que significa que se agruparon *3* moléculas de agua en una partícula DPD. Con este grado de *"coarse-graining"* se ha mostrado que se reproduce la compresibilidad isotérmica del agua a temperatura ambiente (Groot y Warren 1997). Para cada simulación se fijó la cantidad total de partículas DPD libres en $N_T$=4300 variando la concentración de la especie a estudiar para construir la isoterma de adsorción. Se colocaron condiciones periódicas en $x$ y $y$; no así en $z$ para representar el confinamiento. A partir de la simulación se obtuvieron los perfiles de densidad de cada especie y se obtuvo la isoterma de adsorción mediante la integración de estos, mediante la expresión $\Gamma = \int [\rho(z) - \rho b] dz$. En los casos donde todo el polímero que se coloca al inicio se adsorbe, el $\Gamma$ es igual a $\Gamma = \int [\rho(z)] dz$.

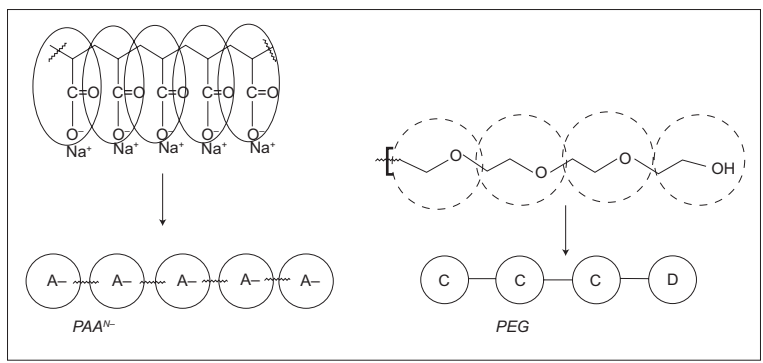

**Fig. 1.** Mapeo del $PAA^{N-}$ y PEG presentes en los sistemas estudiados para las simulaciones DPD

## RESULTADOS Y DISCUSIÓN

### a) Sistema 1: $X_mO_n$ (sustrato)+ $H_2O$ + $PAA^{N-}$ + $(N+r)Na^+$ + $rCl^-$

Se llevó a cabo la simulación DPD electrostática para el **sistema 1**. Para validar la metodología los resultados obtenidos fueron contrastados con los resultados experimentales reportados en la literatura (Huldén y Sjöblom 1990) para la adsorción de la sal de sodio del ácido poliacrílico adsorbido sobre pigmento de $TiO_2$ recubierto con $Al_2O_3$ y $SiO_2$. Los resultados presentados en la literatura, muestran que el recubrimiento del $TiO_2$ con distintos óxidos minerales tiene una fuerte influencia en la adsorción, lo cual se explica por las diferentes características ácido-base de los recubrimientos. Por ejemplo el $TiO_2$ recubierto con $Al_2O_3$ ó ($Al_2O_3$+$SiO_2$) resulta en una superficie básica o ácida, respectivamente. Por ello se observa que el $[PAA^{N-}][Na^+]_N$ se adsorbe en sustrato básico pero no en el ácido. En dicho trabajo (Huldén y Sjöblom 1990) se presentan los resultados de las isotermas de adsorción mostrando este resultado. Aún cuando a este pH la superficie del pigmento básico presenta una ligera carga negativa al igual que el polielectrolito, se produce su adsorción debido a que la repulsión electrostática es más débil que la de otras contribuciones a la energía de adsorción. La ausencia de adsorción del $[PAA^{N-}][Na^+]_N$ sobre superficies ácidas (con carga significativamente negativa) se debe a la fuerte repulsión electrostática presente.

Con base en lo anterior, se llevó a cabo la simulación del $[PAA^{N-}][Na^+]_N$ sobre una superficie básica de óxido metálico del tipo $X_mO_n$, empleando paredes explícitas genéricas para representarlos esencialmente neutras en conjunto pero con cargas puntuales sobre los $n$ átomos de $O$ de $-0.576$ mien-tras que, para los m átomos de X se tiene una carga neta de $+1.152$ equivalente a las cargas parciales de una superficie básica de $TiO_2$ con un recubrimiento de $Al_2O_3$. Estas cargas contribuyen fuertemente a la atracción electrostática con la parte cargada del electrolito, en este caso se incluyó además una interacción DPD repulsora entre los átomos de la pared y los del bulto de 100. Durante la simulación la fuerza iónica se mantuvo constante fijándola con NaCl en una concentración equivalente a 0.1 N.

Con ello se obtuvieron los perfiles de densidad de cada sistema a diferente concentración de $[PAA^{N-}]$ $[Na^+]_N$, las concentraciones manejadas fueron 7, 12, 17, 23 y 28.

La **figura 2** presenta la configuración al equilibrio para distintas concentraciones, el análisis de las configuraciones finales muestra cómo se adsorbe el $PAA^{N-}$ sobre la superficie, el arreglo ordenado de moléculas sugiere el acomodo de éstas sobre la superficie. La **figura 3** presenta los perfiles de densidad obtenidos, éstas gráficas muestran claramente la adsorción del $PAA^{N-}$ en las paredes, indicadas con líneas verticales en los extremos de la gráfica. Observamos además la formación de estructura, a través de la formación de capas subsecuentes lo que también obedece a un ordenamiento sugerido por el empaquetamiento. Integrando los perfiles de densidad es posible obtener la adsorción del sistema $\Gamma$, como se ha discutido en la sección de métodos. En la **figura 4** se muestra la isoterma de adsorción obtenida comparada con los resultados reportados en la literatura (Huldén y Sjöblom 1990). En este caso se nota un comportamiento Langmuir. Con estos datos es posible hacer un análisis del valor del área de saturación de la superficie a través del análisis de la isoterma experimental y por simulación ajustando a un modelo de Langmuir obteniéndose los siguientes resultados para la parte lineal de la isoterma: para el caso experimental un valor de $1/\Gamma_M$=0.882135 mientras que para la simulación $1/\Gamma_M$=0.882944

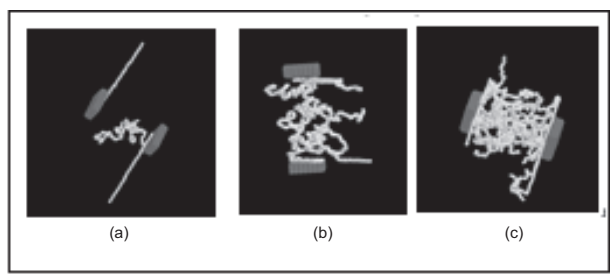

(a)          (b)          (c)

**Fig. 2.** Configuraciones al equilibrio para la adsorción de $PAA^{N-}$ sobre superficie básica de óxido metálico del tipo $X_mO_n$ a distintas concentraciones (a) $[PAA^{N-}]$=4, (b) $[PAA^{N-}]$=17, $[PAA^{N-}]$=28

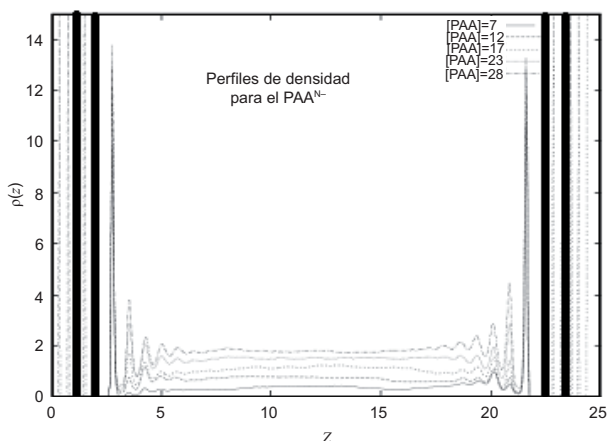

**Fig. 3.** Perfiles de densidad para el $PAA^{N-}$ adsorbido sobre superficie básica de óxido metálico del tipo $X_mO_n$ indicado con líneas gruesas a los extremos

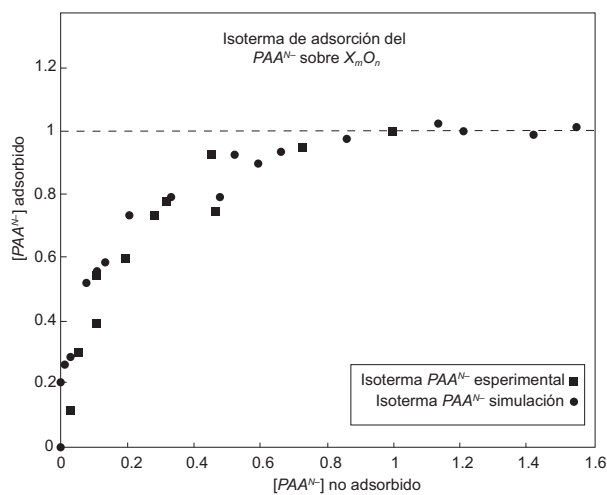

**Fig. 4.** Isotermas de adsorción (experimental y teórica) para el $PAA^- + Na^+$ en $H_2O$ a $[NaCl]=0.1$ N adsorbido sobre superficie básica de óxido metálico del tipo $X_mO_n$

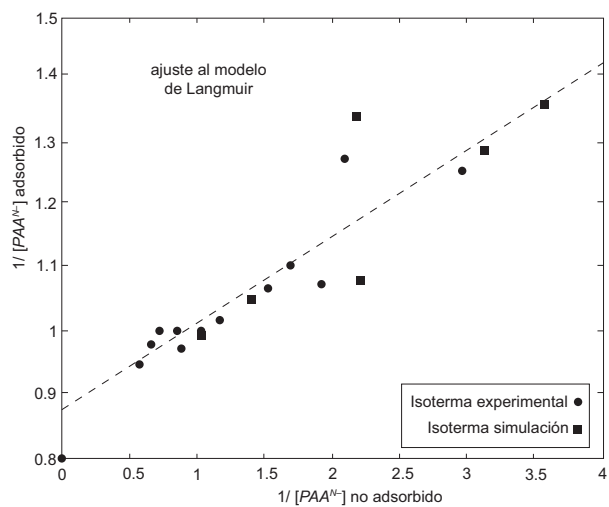

**Fig. 5.** Ajuste tipo Langmuir normalizado de la isoterma de adsorción (experimental y teórica) para el adsorbido sobre superficie básica de óxido metálico del tipo $X_mO_n$

mientras que el valor de $1/\Gamma_M K=0.13694$ para los resultados experimentales y $1/\Gamma_M K=0.13692$ para los resultados obtenidos por simulación. Resulta remarcable la similitud entre ambos resultados. A partir de aquí extraemos el valor de $\Gamma_M=1.13257$ y $K=6.4476$.

La **figura 5** muestra el ajuste a una isoterma tipo Langmuir. Al llevar a cabo la conversión de unidades tomando en cuenta el área superficial del $TiO_2$ recubierto con $(Al_2O_3)$ de 30.22 m²/g y reportado en (Huldén y Sjöblom 1990), se obtiene $\Gamma_M=7.987$ mg de PAA/gr $TiO_2$. El valor reportado del artículo de Huldén es de 6.96 mg de PAA/gr $TiO_2$.

## b) Sistema 2. $X_mO_n$ (sustrato) + $H_2O$ + PEG

El siguiente sistema estudiado fue la adsorción del polietilenglicol (PEG) sobre el mismo sustrato descrito en la sección anterior esto es, sobre dos placas paralelas genéricas representativas de óxidos metálicos $X_mO_n$ con características básicas. Las concentraciones de PEG fueron de 4, 12, 20, 28, 36, 44, 52 y 60. La fuerza iónica se mantuvo en cero es decir no se encuentran presentes iones en este sistema. A partir de los perfiles de densidad se obtuvo la isoterma de adsorción que se presenta en la **figura 6**. En este caso se obtiene que la concentración de saturación $\Gamma_{max}=31$ moléculas de PEG. La **figura 7a** muestra las configuraciones al equilibrio en la concentración de saturación, en las paredes del sustrato (en rojo) se observa la adsorción del PEG representado por cuentas moradas, los extremos con una ligera densidad de carga se esquematizaron con cuentas en gris. La forma en que se satura esta

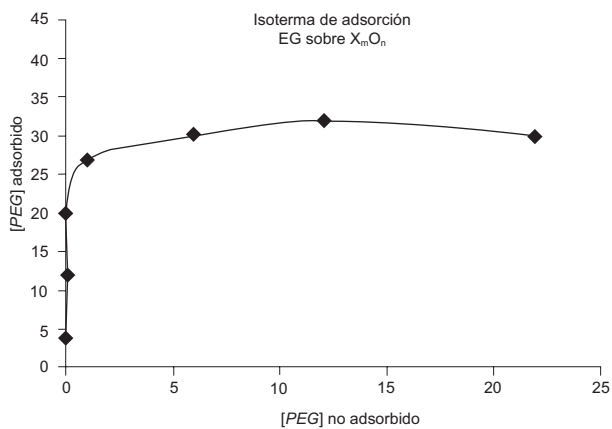

**Fig. 6.** Isoterma de adsorción para el PEG adsorbido sobre superficie básica de óxido metálico del tipo $X_mO_n$

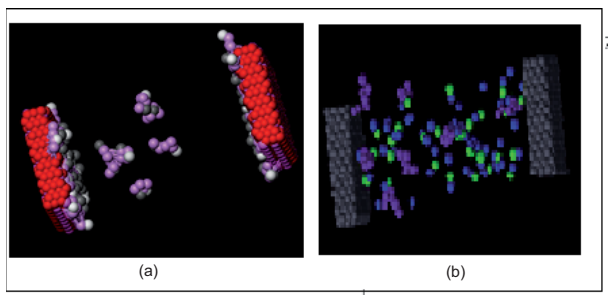

**Fig. 7.** Configuración en la saturación para el sistema 2 (a) y sistema 3 (b). Las moléculas moradas muestran el PEG mientras que las verdes y azules en la figura (b) muestran el ion $UO_2^{2+}$ y su $Cl^{2-}$ respectivamente

superficie es distinta a la anterior ya que la afinidad con el sustrato ácido es distinta (no iónica) y además a que el peso molecular de la molécula es menor observándose una mayor cantidad de moléculas de *PEG* en el bulto.

### c) **Sistema 3**. $X_mO_n$ (sustrato) + $H_2O$ + PEG + $UO_2^{2+}$+$Cl^{2-}$

En este caso se estudió el cambio en el comportamiento de la adsorción del *PEG* sobre la misma superficie de los sistemas anteriores cuando se agrega el ión uranilo y su contraión ($Cl^{2-}$) representado en este caso como una partícula DPD libre con carga $2^+$ y $2^-$. La fuerza iónica total debida a la presencia de este ión se fijó en 0.05 N. La **figura 7b**, muestra la configuración en la saturación para este sistema. En azul y verde se presentan el ion uranilo y su contraión respectivamente; observamos al comparar con el caso anterior (**Fig. 7a**) que los iones uranilo compiten con el PEG (en morado) por la superficie alterando la adsorción de éste sobre el sustrato. La isoterma de adsorción obtenida para el componente PEG se muestra en la **figura 8**, se nota como la presencia del ión uranilo ($UO_2^{2+}$) cambia notablemente el comportamiento al compararlo con la isoterma anterior de PEG sobre el mismo sustrato de la simulación anterior sin iones (ver **Fig. 6**). El ajuste de esta isoterma con un modelo tipo Langmuir muestra un valor de $1/\Gamma_{max} = 0.0670397$ esto es $\Gamma_{max}$=14.916 cuentas DPD de PEG 400 adsorbidos menor que en el caso anterior. Lo anterior sugiere que hay menor adsorción de PEG 400 sobre la superficie en presencia del ión ($UO_2^{2+}$).

### c) **Sistema 4**. $X_mO_n$ (sustrato) + $H_2O$ + [$PAA^{N-}$] $+_N[Na^+]$+ $UO_2^{2+}$+$2Cl^-$

En el cuarto sistema se estudió la adsorción del [$PAA^{N-}$][$Na^+$]$_N$ sobre el mismo sustrato que los sistemas anteriores, en presencia del ión uranilo y su contraión representado de la misma manera que en el sistema 3. La fuerza iónica total debida a la presencia de este ión se fijó nuevamente en 0.05 N. Para cada sistema distinto se llevaron a cabo las simulaciones variando la cantidad de polímero adicionado, la isoterma obtenida se presenta en la **figura 9**. Se observa que bajo estas condiciones de fuerza iónica fijada por el efecto del ión $UO_2^{2+}$ el comportamiento de la adsorción del $PAA^{N-}$ en la isoterma es afectado, al grado de que la forma de la isoterma cambia completamente no siendo posible ajustarla a un comportamiento tipo Langmuir. Es importante notar que en comparación con el sistema 1 que se empleó para la validación de la metodología, la fuerza iónica fijada con NaCl para reproducir los resultados experimentales (0.1 N) es mucho más alta que en el presente caso (0.05 N) originada por la baja concentración del ión $UO_2^{2+}$.

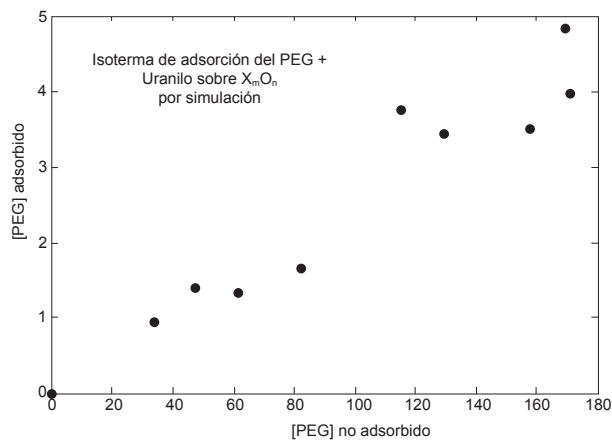

**Fig. 8.** Isotermas de adsorción para el PEG con ion $UO_2^{2+}$ sobre superficie básica de óxido metálico del tipo $X_mO_n$.

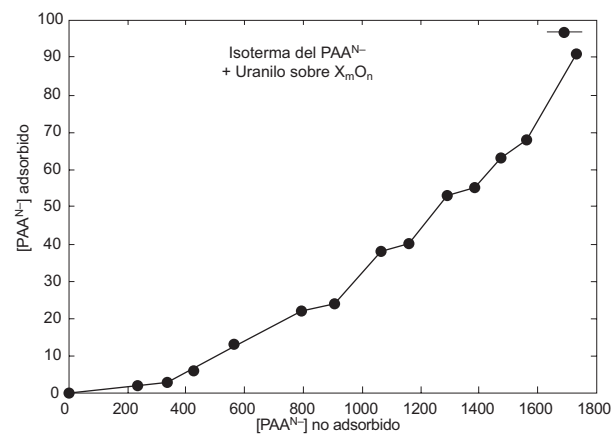

**Fig. 9.** Isotermas de adsorción y configuración en la saturación para el $PAA^{N-}$ + $Na^+$ en $H_2O$ con ion $UO_2^{2+}$ sobre superficie básica de óxido metálico del tipo $X_mO_n$.

## CONCLUSIONES

Los resultados anteriores muestran como, mediante simulaciones DPD electrostáticas es posible obtener las isotermas de adsorción de distintos componentes orgánicos e inorgánicos neutros y cargados, así como estimar la cantidad de material adsorbido por unidad de superficie. La competitividad entre las especies presentes muestra que cuando se agrega un ion uranilo éste compite por la adsorción sobre el sustrato con el resto de los componentes modificándose el equilibrio del sistema simple. Con base en este estudio se concluye que es de suma importancia considerar al sistema con todos los componentes representativos presentes ya que cada una de las moléculas presentes afectarán la constante de adsorción en el sistema y por lo tanto la concentración del contaminante durante el proceso de difusión.

## AGRADECIMIENTOS

## REFERENCIAS

Bear J. (1979), Dynamic of fluids in porous media. 2da. ed. Dover Publications, Nueva York, EUA,764 pp.

Español P. y Warren P. (1995). Statistical mechanics of dissipative particle dynamics. Europhys. Lett. 30, 191-196.

Español P. (2002) Dissipative particle dynamics revisted, SIMU 'Challenges in molecular simulations' newsletter, issue 4, March 2002 (Disponible en http://simu.ulb.ac.be/59-77.)

Esumi K., Nakaie Y., Sakai K. y Torigoe K. (2001). Adsorption of poly(ethyleneglycol) and poly(amidoamine) dendrimer from their mixtures on alumina/water, Colloid. Surface. A 194, 7-12.

Fermeglia M. y Pricl S. (2007). Multiscale modeling for polymer systems of industrial interest. Prog. Org. Coat. 58, 187-199

Fetter C. W. (1994). Applied hydrogeology. 3a. ed., Prentice-Hall, New Jersey, EUA, 691 pp.

Frenkel D. y Smit B. (2002). Understanding molecular simulation: from algorithms to applications. 2a ed. Academic Press, San Diego, EUA  460 pp.

González-Melchor M., Mayoral E., Velázquez M.E. y Alejandre J. (2006). Electrostatic interactions in dissipative particle dynamics using the Ewald sums. J. Chem. Phys. 125, 1-8.

Groot D. y Warren P. B. (1997), Dissipative particle dynamics: bridging the gap between atomistic and mesoscopic simulation. J. Chem. Phys. 107, 4423-4435.

Groot D. (2003), Electrostatic interactions in dissipative particle dynamics-simulation of polyelectrolytes and anionic surfactants. J. Chem. Phys. 118, 11265-11277.

Hoogerbrugge P. J. and Koelman J. M. V. A. (1992), Simulating microscopic hydrodynamic phenomena with dissipative particle dynamics. Europhys. Lett., 19, 155-160.

Huldén M., and Sjöblom E. (1990), Adsorption of some common surfactants and polymers on TiO2-pigments. Progr. Colloid. Polym. Sci. 82, 28-37

IAEA (1999) No. WS-G-1.1 Safety Guide. Safety assessment for near surface disposal of radioactive waste. International Atomic Energy Agency. Safety standards series. Vienna, 42 pp .

Iturbe J.L. (2001), Fundamentos de radioquímica, Universidad Autónoma del Estado de México, Estado de México, México, 406 pp.

SENER (1996). Norma Oficial Mexicana NOM-022/1-NUCL-1996, Requerimientos para una instalación para el almacenamiento definitivo de desechos radiactivos de nivel bajo cerca de la superficie. Parte 1, Sitio. Secretaría de Energía. Comisión Nacional de Seguridad Nuclear y Salvaguardias. Diario Oficial de la Federación. 5 de Septiembre de 1997.

Toso J.P and Velasco R. H. J. (2001), Describing the observed vertical transport of radiocesium in specific soils with three time-dependent models. J. Environ. Radioactiv. 53, 133-144.